\def\be{\begin{equation}}

\def\ee{\end{equation}}

\def\ba{\begin{eqnarray}}

\def\ea{\end{eqnarray}}

\def\n{\nonumber \\}

\def\b{\bibitem}

\documentclass[12pt]{article}

\begin{document}

\title{Physical meaning of two-particle
HBT measurements in case of  correlated emission}

\author{A.Bialas and K.Zalewski  \\ M.Smoluchowski Institute of Physics
\\Jagellonian University, Cracow\thanks{Address: Reymonta 4, 30-059
Krakow, Poland; e-mail:bialas@th.if.uj.edu.pl;}\\and \\
Institute of Nuclear Physics PAN, Cracow} \maketitle

\begin{abstract} It is shown that, in the presence of correlations in
particle emission, the measured HBT radii are related to the
correlation range rather than to the size of the interaction
volume. Only in the case of weak correlations the standard
interpretation may be applicable. The earlier discussion
\cite{prdr} of the short-range correlations in configuration space
is generalized to include also the correlations of particle
momenta.
\end{abstract}

{\bf 1.} Measurements of HBT correlations in multiparticle
production provide important information on the production
mechanism, in particular on the space-time structure of the
particle emission region \cite{rev}. To obtain this information,
however, it is necessary to rely on some specific theoretical
interpretation of the observed phenomena. The results are model
dependent: The physical meaning assigned to the measured
quantities does depend on the theoretical input.

In the standard treatment of this problem one usually starts with
a model where particles are uncorrelated (except for Bose-Einstein
correlations) and then corrects the results by including final
state interactions. This includes corrections for Coulomb
interactions, low energy particle interaction parametrized by
scattering lengths and effects of resonances \cite{kk}. In the
present paper we discuss correlations due to strong interactions
in the production process. Some such correlations are known to
occur \cite{foa}, some others, still hypothetical, may be -
hopefully - uncovered by the HBT measurements \cite{prdr}.

To simplify the presentation we consider only the two-dimensional
(transverse) distributions\footnote{i.e. distributions integrated
over some interval of the longitudinal variables.}, taken as
Gaussians to avoid complicated integrations which only obscure the
essential points of our argument. In this case Wigner functions
$W({\bf p}_1,...{\bf p}_n;{\bf x}_1,...{\bf x}_n)$ can be used
instead of the more complicated emission functions
$S(p_1,...p_n;x_1,...x_n)$. The Wigner functions are real
functions of momenta and positions and are in a well-defined sense
\cite{W} the best quantum analog of particle density in
phase-space. Therefore the parameters characterizing the Wigner
functions can be interpreted\footnote{Given all the caveats
related to the fact that we are dealing with quantum phenomena
\cite{rev, z}.} as the parameters characterizing the space
distribution of sources and their momentum spectra \cite{spr}.

The density matrix in momentum space is related to the Wigner
function by  the formula:

\ba \rho({\bf p}_1,...,{\bf p}_n;{\bf p}_1',...,{\bf p}_n')=
\n=\int d^2x_1...d^2x_n \exp\left[i\left({\bf Q}_1{\bf
x}_1+...+{\bf Q}_n{\bf x}_n\right)\right] W({\bf K}_1,...,{\bf
K}_n;{\bf x}_1,...{\bf x}_n) \label{1z} \ea
where $\;{\bf K}_i= (
{\bf p}_i+{\bf p}'_i)/2$ and $ {\bf Q}_i= {\bf p}_i-{\bf p}_i'.
$

It follows that the momentum distribution of particles can be
expressed as

\ba \Omega_0({\bf p}_1,...,{\bf p}_n)= \rho({\bf p}_1,...,{\bf
p}_n;{\bf p}_1,...,{\bf p}_n)=\n= \int d^2x_1...d^2x_n W({\bf
p}_1,...,{\bf p}_n;{\bf x}_1,...{\bf x}_n)    \label{3z} \ea
Similarly, for the coordinate distribution we have

\ba \Omega_0({\bf x}_1,...,{\bf x}_n)= \rho({\bf x}_1,...,{\bf
x}_n;{\bf x}_1,...,{\bf x}_n)=\n= \int d^2p_1...d^2p_n W({\bf
p}_1,...,{\bf p}_n;{\bf x}_1,...{\bf x}_n)    \label{3zz} \ea

For the momentum distribution of identical bosons we have to
symmetrize  the production amplitudes. This modifies the momentum
distribution (see, e.g., \cite{bk}) into

\ba \Omega({\bf p}_1,...,{\bf p}_n)=\frac1{n!} \sum_{P,P'}
\rho({\bf p}_{i_1},...,{\bf p}_{i_n};{\bf p}_{i'_1} ,...,{\bf
p}_{i'_n})    \label{4z} \ea where the sum runs over all
permutations $P$ and $P'$ of $(i_1,...i_n)$ and
$(i'_1,...i'_n)$.\footnote{For fermions there is an extra minus
sign when $P$ and$P'$ are odd with respect to each other.} This is
the key formula which explains the main interest in the HBT
measurements: the distribution of identical particles opens a
window to the non-diagonal elements of the density matrix and thus
also to the Wigner function. It is also clear, however, that this
information is not sufficient to obtain full information about the
distribution of sources. Thus further theoretical input is needed.

The purpose of the present paper is to discuss the physical
meaning of the measured two-particle HBT parameters in terms of
the characteristics of the momentum and coordinate distribution of
the sources as described by the Wigner function. The well-known
case of uncorrelated emission (for recent reviews, see e..g.
\cite{rev}) is summarized briefly in the next section. The
emission of particles correlated in pairs is described in Section
3. In Section 4 a more realistic situation, when only a fraction
of the particles is emitted in pairs while others remain
uncorrelated, is considered. The experimental consequences are
discussed in Sections 5 and 6. Our conclusions are listed in the
last section.

{\bf 2.} The assumption of uncorrelated production means that the
Wigner function factorizes into a product of single particle
Wigner functions. Of course this factorization is then satisfied
also for the unsymmetrized density matrix.

To illustrate the consequences of this Ansatz and to fix our
notation, consider a single particle Wigner function in the most
general Gaussian form\footnote{ As already mentioned in the
Introduction, all vectors are two-dimensional.}$^,$\footnote{This
model is sometimes referred to as the Zajc model \cite{zajc}.}

\ba W({\bf p},{\bf x})=\frac1{4\pi^2\Delta_u^2(R_u^2-r_u^2)}
\exp\left[-\frac{{\bf p}^2} {2\Delta_u^2}-\frac{({\bf x}-r_u{\bf
p}/\Delta_u)^2}{2(R_u^2-r_u^2)}\right]\label{6z} \ea One sees that
the parameter $r_u$ is responsible for momentum-position
correlation. From (\ref{6z}), using (\ref{3z}) and (\ref{3zz}), we
derive for single particle distributions

\ba \Omega_0({\bf p})= \int d^2x W({\bf p},{\bf x})= \frac1{2\pi
\Delta_u^2}\exp\left[-\frac{{\bf
p}^2}{2\Delta_u^2}\right];\;\;\;\n \Omega_0({\bf x})= \int d^2p
W({\bf p},{\bf x}) =\frac1{2\pi R_u^2} \exp\left[-\frac{{\bf
x}^2}{2R_u^2}\right].  \label{6zz} \ea

One sees that the parameter $\Delta_u$ describes the width of the
distribution in momentum space whereas $R_u$ determines the size
of the system in configuration space.

Using (\ref{1z}) and (\ref{4z}) , we  obtain the two-particle
distribution for  identical particles:

\ba \Omega({\bf p}_1,{\bf p}_2)=\frac1{4\pi^2\Delta_u^4}
\exp\left[-\frac{{\bf p}_1^2+{\bf p}_2^2}{2\Delta_u^2}\right]
\left\{1+\exp\left[-({\bf p}_1-{\bf
p}_2)^2R_{HBT}^2\right]\right\}           \label{8z} \ea
where

\ba R_{HBT}^2\equiv R_u^2-r_u^2 -\frac1{4\Delta_u^2}  \label{9z}
\ea

One sees that in this simple case measurements of the single
particle distribution and pair distribution allow to determine
$\Delta_u$ and $R_{HBT}$. One also sees from (\ref{9z}) that these
two parameters are not sufficient to determine $R_u$, the size of
the system in configuration space \cite{sib}. To this end it is
necessary to know the correlation between the momentum and the
position of the emission point of a particle, as expressed by the
parameter $r_u$.

{\bf 3.} The most general Gaussian two-particle Wigner function,
symmetric with respect to simultaneous exchange of the particle
momenta and positions, can be written as

\ba W_c({\bf p}_1,{\bf p}_2;{\bf x}_1,{\bf x}_2)=
\frac1{16\pi^4\Delta_+^2\Delta_-^2(R_+^2-r_+^2)(R_-^2-r_-^2)}\n
\exp\left[-\frac{{\bf p}_+^2}{\Delta_+^2}-\frac{{\bf
p}_-^2}{\Delta_-^2}\right] \exp\left[-\frac{({\bf x}_+-r_+{\bf
p}_+/\Delta_+)^2}{R_+^2-r_+^2} -\frac{({\bf x}_--r_-{\bf
p}_-/\Delta_-)^2}{R_-^2-r_-^2}\right]  \label{5} \ea where $\;{\bf
p}_{\pm}= ({\bf p}_1\pm {\bf p}_2)/2\;$ and  $\;{\bf x}_{\pm}=
({\bf x}_1\pm {\bf x}_2)/2$. Note that if

\ba \Delta_- =\Delta_+;\;\;\;R_+=R_-;\;\;\;r_+=r_-  \label{6a} \ea
the Wigner function factorizes and the problem reduces to the one
discussed in the previous section.

One sees from (\ref{5}) that $r_{\pm}$ are responsible for the
correlations between positions and momenta. To see the physical
meaning of the other 4 parameters we calculate the distribution of
momenta

\ba \Omega_0({\bf p}_1,{\bf p}_2)=
\frac1{4\pi^2\Delta_+^2\Delta_-^2} \exp\left[-\frac{{\bf
p}_1^2+{\bf p}_2^2}{2\Delta_+^2} -\frac{({\bf p}_1-{\bf
p}_2)^2}{2\omega^2}\right] \label{5ab} \ea
and positions

\ba \Omega_0({\bf x}_1,{\bf x}_2)= \frac1{4\pi^2R_+^2R_-^2}
\exp\left[-\frac{{\bf x}_1^2+{\bf x}_2^2}{2R_+^2}- \frac{({\bf
x}_1-{\bf x}_2)^2}{2\xi^2}\right]  \label{5b} \ea
where

\ba \frac1{\omega^2} =\frac1{2\Delta_-^2}-
\frac1{2\Delta_+^2};\;\;\; \frac1{\xi^2}=
\frac1{2R_-^2}-\frac1{2R_+^2}.   \label{5c} \ea
From this we see
that $\Delta_+^2$ describes the momentum distribution, whereas
$\omega^2$ describes the correlations between the momenta in the
pair. Similarly, $R_+^2$ describes the distribution of the
particle positions while $\xi^2$ describes correlations between
the positions of particles in the pair. Note that $\omega^2$ and
$\xi^2$ are not necessarily positive. Note also that correlations
do indeed disappear ($1/\omega=1/\xi=0$) when condition (\ref{6a})
is satisfied.

Using (\ref{5}), (\ref{1z}) and (\ref{4z}), the two-particle
density matrix is obtained:

\ba \Omega({\bf p}_1,{\bf p}_2)= \Omega_0({\bf p}_1,{\bf
p}_2)\left(1+ \exp\left[-({\bf p}_1-{\bf p}_2)^2
R_c^2\right]\right) \label{4} \ea where $\Omega_0$ is given by
(\ref{5ab})
and

\ba R_c^2=R_-^2-r_-^2-1/4\Delta_-^2 .  \label{3} \ea
One sees that
$\Omega({\bf p}_1,{\bf p}_2)$  depends only on three parameters:
$\Delta_+^2, \Delta_-^2$,  and $R_-^2-r_-^2$, whereas $R_+^2$ and
$r_+^2$   do not have any impact on the momentum distribution.

Using (\ref{5c}) we obtain

\ba R_-^2= \frac {\xi^2R_+^2}{\xi^2+2R_+^2} \label{5ac} \ea
which
explicitely shows the effect of correlations in configuration
space on the physical interpretation of the HBT measurements.

Note that for positive correlations ($\xi^2>0$) $R_-^2$ is always
smaller than both $\xi^2/2$ and $R_+^2$. In particular, when
$\xi^2\ll R_+^2$ we have $R_-^2 \approx \xi^2/2$. In this case the
HBT measurements give only information on correlations and {\it
not} on the size of the system in configuration space.

One also sees that for negative correlations $R_-^2 $ is always
greater than $R_+^2$.

This discussion shows that correlations in configuration space can
significatly influence the interpretation of the measured HBT
parameters. Only if there are no correlations ($1/\xi^2 = 0$),
$R_+$ and $R_-$ are identical and by this "accident" one can
obtain information about the total volume of the reaction

{\bf 4.} In the previous section we have discussed the situation
when {\it all} pairs of the emitted particles are correlated. This
is an interesting theoretical exercise which, however, hardly
corresponds to reality. The measured HBT correlations indicate
that the data are in reasonable agreement with the assumption of
uncorrelated production. This suggests that to discuss practical
consequences of our formalism it is more appropriate to consider a
situation when correlated emission affects only a fraction of all
the particles, the others remaining uncorrelated.

The formalism developped in Sections 2 and  3 is well suited to
cover this case. We  write the Wigner function as a sum of two
terms: One describing the uncorrelated emission and the other
responsible for the correlations. Following the discussion of
sections 2 and 3  we  write

\ba W({\bf p}_1,{\bf p}_2;{\bf x}_1,{\bf x}_2)= w_uW_u({\bf
p}_1,{\bf x}_1)W_u({\bf p}_2,{\bf x}_2) +w_cW_c({\bf p}_1,{\bf
p}_2;{\bf x}_1,{\bf x}_2) \label{i} \ea
where $W_u({\bf p},{\bf
x})$ is given by (\ref{6z}) and $W_c({\bf p}_1,{\bf p}_2;{\bf
x}_1,{\bf x}_2)$ by (\ref{5}). $w_u$ is the probability that the
considered particles are uncorrelated and $w_c=1-w_u$ is the
probability that they were emitted as a correlated pair.

The density matrix is thus given by a sum of two terms, one
constructed from $W_u$ and the other from $W_c$. This gives the
single particle momentum distribution\footnote{In (\ref{om0}) the
corrections due to BE correlations are neglected. They are
expected to be small at high energies.}

\ba \Omega_0({\bf p}_1)= \frac{1}{2\pi\Delta_u^2}e^{-{\bf
p}_1^2/2\Delta_u^2} \Phi_0({\bf p}_1) \label{om0} \ea
where

\ba \Phi_0({\bf p}_1)= w_u+ w_c
\frac{2\Delta_u^2}{\Delta_+^2+\Delta_-^2} e^{-{\bf
p}_1^2/\eta^2}\label{bvi} \ea
represents the modification of the
single particle spectrum due to the correlated emission. Here

\ba \frac1{\eta^2}=
\frac1{\Delta_+^2+\Delta_-^2}-\frac1{2\Delta_u^2}.    \label{avii}
\ea

Using (\ref{i}) and employing (\ref{1z}) and (\ref{4z}), the
momentum distribution for identical particles $\Omega({\bf
p}_1,{\bf p}_2)$ can now be derived   and thus one can construct
the usually measured quantity

\ba C({\bf p}_1,{\bf p}_2)\equiv \frac{\Omega({\bf p}_1,{\bf
p}_2)} {\Omega_0({\bf p}_1)\Omega_0({\bf p}_2)}   \label{avi} \ea
where $\Omega_0({\bf p}_1)$ is the single-particle distribution
{\it in the events whith at least one pair of identical
particles}, given by (\ref{om0}). The result is

\ba C({\bf p}_1,{\bf p}_2)= w_u C_u({\bf p}_1,{\bf p}_2)
+w_cC_c({\bf p}_1,{\bf p}_2)  \label{xixi} \ea
with

\ba C_u({\bf p}_1,{\bf p}_2)= \frac{1+e^{-({\bf p}_1-{\bf
p}_2)^2R_{HBT}^2}}{\Phi_0({\bf p}_1)\Phi_0({\bf p}_2)}
\label{xiyi} \ea
and

\ba C_c({\bf p}_1,{\bf p}_2)=
\frac{\Delta_u^4}{\Delta_+^2\Delta_-^2} \frac{e^{-({\bf p}_1+{\bf
p}_2)^2/2\chi_+^2}e^{-({\bf p}_1-{\bf
p}_2)^2/2\chi_-^2}}{\Phi_0({\bf p}_1)\Phi_0({\bf p}_2)}
\left[1+e^{-({\bf p}_1-{\bf p}_2)^2R_c^2}\right]\label{fic1} \ea
with

\ba \frac1{\chi_{\pm}^2}=
\frac1{2\Delta_{\pm}^2}-\frac1{2\Delta_u^2};\;\;\; \label{chi} \ea

{\bf 5.} The formulae (\ref{xixi})-(\ref{fic1}) describe the HBT
measurements for a general superposition of uncorrelated and
correlated emission. They thus cover a wide range of possible
physical situations.

To discuss their interpretation we have to consider the possible
origin of these two contributions. The uncorrelated emission may
stem either from directly produced pions or from the pions emitted
from uncorrelated clusters (resonances). The correlated emission
may reflect (i) a genuine structure of the source \cite{prdr} or
(ii) the interaction between pions. The attractive interactions
lead to positive correlations ($\xi^2>0$). They are usually
represented as clusters of pions. The repulsive interactions
(which were never observed\footnote{As already stated in Section
1, we discuss here only correlations due to strong interactions in
the production process.}) would give negative correlations
($\xi^2<0)$.

As seen from (\ref{xixi})-(\ref{fic1}), for positive correlations
one may expect the two components, $C_u$ and $C_c$, to have
different ranges in $({\bf p}_1-{\bf p}_2)^2$. The difference may
be large, especially in heavy ion collisions. Indeed, in this case
the range of the first one ($\sim 1/R_{HBT}^{2}$) is determined by
the size of the whole system, whereas the range of the second one
($\sim 1/R_c^2$) is determined by the geometrical size of clusters
(and/or of local fluctuations) and by the momentum distributions.

We shall consider in detail the  generic scenario when {\it all}
particles are emitted from uncorrelated sources \cite{prdr}.   The
single particle distribution is then fully determined by the
distribution and decay properties of the emitting sources. The
condition

\ba \int d^2x_2 d^2p_2 W_c({\bf p}_1,{\bf p}_2;{\bf x}_1,{\bf
x}_2) = W_u({\bf p}_1,{\bf x}_1)   \label{55i} \ea
implies

\ba 2\Delta_u^2=\Delta_+^2+\Delta_-^2;\;\;2R_u^2=R_+^2+R_-^2;\;\;
2r_u\Delta_u=r_-\Delta_-+r_+\Delta_+       \label{5i} \ea and,
naturally,  $\Phi_0({\bf p}) \equiv 1$.

A special case of this scenario (particle emission from
independent granules) was discussed in \cite{prdr} where it was
furthermore assumed that (i) the distribution of sources is
momentum-independent ($1/\Delta_+^{2}=0$) and (ii) the momentum
dependence in source decay may be neglected with respect to
dependence on difference of momenta ($1/\Delta_-^{2}\ll R_c^2$,
$R_{HBT}^2$). Under these conditions\footnote{They are too
restrictive: to obtain (\ref{xixy}) it is enough to assume
$\Delta_+=\Delta_+=\Delta_u$, i.e., no correlations in momentum
space.} the expression for the correlation function considerably
simplifies

\ba C({\bf p}_1,{\bf p}_2)= 1+ w_u e^{-({\bf p}_1-{\bf
p}_2)^2R_{HBT}^2} + w_c\frac{\Delta_u^4}{\Delta_+^2\Delta_-^2}
e^{-({\bf p}_1-{\bf p}_2)^2R_c^2} \label{xixy} \ea
where $w_c
=1/n$ and $n$ is the total number of sources.

One sees clearly the two-component structure of the correlation
function\footnote{A sum of two Gaussians in the two-particle
correlation function was also considered for another reason in
\cite{lpo}.}. As pointed out in \cite{prdr}, the observation of
the second term may serve as an indication of the clustering
and/or of the granular structure of the emission region in heavy
ion collisions. The size of the granules (clusters) may be read
off from the range of the second component.

The simple formula (\ref{xixy}) illustrates very well the basic
physics of the problem. As seen from our general expression
(\ref{fic1}), however, the actual shape of the second component
may be significantly influenced by the momentum dependence of the
emitting sources. It is true that $1/\Delta_+^2$ and
$1/\Delta_-^2$, being of the order of 1 fermi$^2$ or less, are
small as compared to $R_{HBT}^2$ which (in heavy ion collisions)
is  of the order of (several fermi)$^2$. They may well be
comparable, however, with $R_c^2$ which need not be much larger
than 1 fermi$^2$. Thus neglecting the momentum dependence of the
emitting sources \cite{prdr} may be a too drastic simplification.

Moreover, even in absence of the correlations in configuration
space (i.e., for $R_+=R_-; r_+=r_-=0$) the two component structure
of the correlation function persists. Indeed, we obtain from
(\ref{xixi})-(\ref{fic1})

\ba C({\bf p}_1,{\bf p}_2)= w_u +
\left(w_u+w_c\frac{\Delta_u^4}{\Delta_+^2\Delta_-^2} e^{-({\bf
p}_1+{\bf p}_2)^2/2\chi_+^2}\right)e^{-({\bf p}_1-{\bf
p}_2)^2R_{HBT}^2} +\n+ w_c\frac{\Delta_u^4}{\Delta_+^2\Delta_-^2}
e^{-({\bf p}_1+{\bf p}_2)^2/2\chi_+^2}e^{-({\bf p}_1-{\bf
p}_2)^2/2\chi_-^2}    \label{nn} \ea
The two-component structure
is recovered but now the momentum correlations and not the
correlations in configuration space are responsible for it.

We conclude that, although the two-component structure of the HBT
measurements seems a robust consequence of the correlated
emission, the physical meaning of the measured parameters is by no
means unique. Thus we feel that in the analysis of actual
experiments our general approach, summarized in the formulae
(\ref{xixi})-(\ref{fic1}), may be needed to account for the
observations and to give the correct physical meaning to the
measured parameters.

{\bf 6.}  Several comments are in order.

(i) One may  note that, since for positive correlations one
naturally expects $\Delta_+^2>\Delta_-^2$, (\ref{5i}) implies that
$\chi_-^2 >0$ and $\chi_+^2 <0$. This means that $C_c$ (c.f.
(\ref{fic1})) increases with increasing momentum of the pair. This
effect may turn out helpful for identification of the second
component\footnote{This conclusion relies heavily on the condition
(\ref{5i}) and thus needs not be generally valid.}.

(ii) It is worth to remember that there are several reasons why
the conditions (\ref{5i}), relating the correlated and the
uncorrelated distributions, may be violated (also the probability
$w_c$ of correlated emission need not be equal to $1/n$). First,
not {\it all} particles are emitted in clusters, some of them are
produced directly. Second, most of the clusters observed in
hadronic collisions are characterized by fairly small multiplicity
(about three particles on the average) and rather small charge
\cite{foa}. Therefore only a small fraction of all clusters emit
{\it two} identical charged pions and there is no obvious reason
why they should have the same properties as an average cluster.
Thus although one may hope that the discussion of the previous
section describes correctly the basic physics of the problem, the
quantitative analysis may require the more flexible approach.

(iii) Finally, let us comment on the possibility of {\it negative}
correlations, i.e. repulsive interaction ($\xi^2 <0$,
$\omega^2<0$). In this case the cluster picture is not applicable.
From (\ref{5ac}) we deduce $R_- >R_+$. Since $R_+$ is expected to
be close to $R_u$, we conclude that $R_c > R_{HBT}$, i.e., the
range of the second component is {\it shorter} than that of the
first one. Thus an observation of an abnormally narrow peak in the
distribution of $ ({\bf p}_1-{\bf p}_2)^2$ may be an indication of
repulsive interactions in the system. It would be interesting to
analyze the data keeping this perspective in mind\footnote{Another
well-known reason for such a narrow peak is the presence of the
long-living resonances \cite{gras}.}.

{\bf 7.} In conclusion, we have analyzed the effects of
interparticle corelations in particle emission on the measurements
of quantum interference. It has been shown that the physical
interpretation of the measured parameters is significantly
influenced by the presence of such correlations. In particular,
for strongly correlated systems the measured range of the HBT
effect is related to the correlation range rather than to the size
of the interaction volume. Only in the case of weak correlations
the standard interpretation may be applicable. The short-range
positive correlations in configuration space were discussed in
detail. The analysis given in \cite{prdr} was generalized. A
possibility to uncover negative interparticle correlations, if
any,  was pointed out.

\vspace{0.3cm}

{\bf Acknowledgements}

\vspace{0.3cm}

This investigation was supported in part by the Grant No 2 P03B
09322 of Polish Commitee for Scientific Research.

\end{document}